\documentclass[12pt, english]{article}

\usepackage[a4paper, margin=2.5cm,footskip=2cm]{geometry}
\usepackage{times}
\usepackage[utf8x]{inputenc}
\usepackage[T1]{fontenc}
\usepackage{setspace}
\usepackage{mathtools}
\usepackage{float}
\usepackage{graphicx}
\usepackage[authoryear]{natbib}
\setcitestyle{authoryear,open={(},close={)}}
\usepackage{color}
\usepackage{microtype}
\usepackage[ruled,vlined]{algorithm2e}
\usepackage{hyperref}
\usepackage[autostyle]{csquotes}
\usepackage{booktabs}
\usepackage{longtable}
\usepackage{tabulary}

\usepackage{tikz}
\usetikzlibrary{fit,positioning,arrows.meta, shapes, chains}
\tikzset{neuron/.style={shape=circle, minimum size=0.9cm, 
  inner sep=0, draw, font=\small}, io/.style={neuron, fill=gray!20}}
\usetikzlibrary{decorations.pathreplacing}
\usetikzlibrary{fadings}

\definecolor{blue}{RGB}{41,5,195}
\hypersetup{colorlinks=true, allcolors=blue}

\title{A Systematic Comparison of Forecasting for Gross Domestic Product in an Emergent Economy} 

\author{
Kleyton da Costa \thanks{kleyton.vsc@gmail.com}\\
Felipe Leite Coelho da Silva \thanks{Department of Mathematics at Federal Rural University of Rio de Janeiro}\\
Josiane da Silva Cordeiro \thanks{Department of Mathematics at Federal Rural University of Rio de Janeiro}\\
André de Melo Modenesi \thanks{Institute of Economics at Federal University of Rio de Janeiro}
\vspace{0.5cm}
}

\date{\today}

\begin{document}

\maketitle

\begin{abstract}
    \noindent Gross domestic product (GDP) is an important economic indicator that aggregates useful information to assist economic agents and policymakers in their decision-making process. In this context, GDP forecasting becomes a powerful decision optimization tool in several areas. In order to contribute in this direction, we investigated the efficiency of classical time series models, the state-space models, and the neural network models, applied to Brazilian gross domestic product. The models used were: a Seasonal Autoregressive Integrated Moving Average (SARIMA) and a Holt-Winters method, which are classical time series models; the dynamic linear model, a state-space model; and neural network autoregression and the multilayer perceptron, artificial neural network models. Based on statistical metrics of model comparison, the multilayer perceptron presented the best in-sample and out-sample forecasting performance for the analyzed period, also incorporating the growth rate structure significantly. 
\end{abstract}





\pagebreak

\section{Introduction}

The economic activity of a country can be influenced by several factors that subject economic agents to change their consumption and investment decisions, in addition to impacting other results, such as inflation and unemployment. Such factors, or shocks, resulting  from the modification of economic policies, in the level of production technology, through meteorological changes etc. The gross domestic product (GDP) is one of the main indexes for measuring the level of economic activity, and the forecast of its trajectory provides useful information concerning the future economic trend in the short term, acting as an object for the expectation of economic behavior.

Significant impacts on economic activity arise through crises. They are a dysfunction inherent in the free market system. Through the development of information transmission technologies and the global integration of markets, the scope and frequency of these dysfunctions have been expanded. Beginning in the second quarter of 2014, the Brazilian economic crisis is still the subject of many analyzes, with no consensus on the generating variables, as well as their consequences. In the second quarter of 2016, the GDP growth rate accumulated in four quarters had reached the lowest level of the last two decades (-4.6 \%). The data show that the recovery (after a significant drop) was not complete, followed by a period of stagnation in the country's growth rate. \cite{paula2017} analyzed the ineffectiveness of counter-cyclical policies - between 2011 and 2014 - as a result of problems in the coordination of macroeconomic policy; and also by the occurrence of exogenous shocks, such as the deterioration of trade terms and the water crisis that occur in period. \cite{filho2017} argues that the origin of the Brazilian economic crisis was due to a series of supply and demand shocks that (mostly) were caused by wrong public policies, contributing to the reduction of growth potential in the Brazilian economy and to the increase in tax cost.

According to \cite{feijo2013}, the most relevant aggregates that derive from the System of National Accounts are the measures of product, income and expenditure. The Macroeconomic Aggregates are statistical constructions that synthesize the productive effort of a given country or region and its possible consequences on the generation of income and expenditure for a specific period of time. By definition, the GDP of a country or region represents production\footnote{The socially organized economic activity that aims to create goods and services to be traded on the market and/or they are achieved by means of factors production (land, capital and labor) traded on the market\citep{ibge2016}.} of all production units of the economy - government, self-employed workers, companies etc. - in a given period, usually quarterly or annually, at market prices\footnote{Economic transactions with observed or imputed market value.}.

\cite{blanchard2017macroeconomics} presents two ways of interpreting GDP. The so-called nominal GDP is defined as the sum of quantities of final goods multiplied by the current price of the goods, that is, considering the inflationary effect during the calculation period. Real GDP takes into account constant prices and sets a given year as a base, excluding the effect of price increases.

Restricting itself to GDP as an instrument for efficiently measuring the quality of life of the population has theoretical and practical limitations. The growth in production is not a sufficient condition for improving well-being (education, health, culture, security etc.). This is because the quality of economic growth is not part of the scope defined for the calculation of GDP. There is the possibility of an expansion sustained by war expenditures (production of supplies and weapons, construction of military installations etc.) or through the reconstruction of a region that has been affected by natural disasters (hurricanes, earthquakes, floods etc.), that it is reasonable to understand themselves as issues that do not promote economic and social well-being or are motivated to do so. Angus Deaton, the Economics Nobel Prize in 2015 says,

\begin{quotation} 
If crime goes up, and we spend more on prisons, GDP will be higher. If we neglect climate change, and spend more and more on cleaning up and repairing after storms, GDP will go up, not down; we count the repairs but ignore the destruction. \citep{deaton2013great}
\end{quotation}

Time series analysis has proven to be an effective tool to understand the behavior patterns of a dataset distributed sequentially over time, with a wide range of models for the purpose of analyzing and predicting trends and seasonality. Seasonal Autoregressive Integrated Moving Average (SARIMA) and Holt-Winters method are considered classical time series models and the class of dynamic linear models is part of the Bayesian approach. Regarding the contributions that use classic models, the analysis constructed by \cite{abozanel2019} for Egypt's annual GDP between the years 1965 and 2016, with forecasting of ten years ahead (2017 to 2026), presented results that pointed to the country's GDP growth during the period under analysis; \cite{wabomba2016} estimated Kenya's GDP between 2013 and in 2017. The result obtained was significant growth in the Kenyan economy in the period; \cite{agrawal2018} modeled the series of India's real GDP growth rate from 1996 to 2017. In the analyzed data, the ARIMA model did not show any more significant results than other models. The author also used the Holt-Winters model and linear trend, both showing similar results each other, and \cite{waslley2020} found significant results using ARIMAX and SARIMAX models (take into account exogenous variables) for the forecast of Brazilian annual and quarterly real GDP for the year 2019.  

For the Bayesian approach and the class of state-space models, \cite{piccoli2015} analyzed four dynamic linear models to identify the one with the best forecasting capacity for nominal GDP in the United States. Best results were obtained using a multivariate model SUTSE (Seemingly Unrelated Time Series Equations) that considered as variables the nominal GDP, the industry production index, the consumer price index (inflation), and the quarterly interest rate for US Treasury bills; \cite{rees2015} built new measures for Australia's GDP growth, using state-space methods. The results found have a high correlation with the figures published officially for GDP growth. However, the measures are less volatile, easier to predict, and achieved good results in nowcasting; \cite{issler2016} estimate Brazilian real monthly GDP with state-space representation and also find good results in forecasting when compared with Central Bank Economic Activity Index (IBC-Br)\footnote{Monthly indicator of national economic activity published by Central Bank of Brazil.}; \cite{migon1993} developed a study about the performance of Bayesian Dynamic Models applied to a set of Brazilian macroeconomics time series (industrial productivity index,  the balance of trade, components of GDP and others) between the period 1970 to 1990. The comparison was made between the dynamic models and  classical structured models and obtained results indicate that the Bayesian approach was similar to the classical approach. Another applied study was developed by \cite{baurle2020gdp}, with the aim of forecasting GDP in the euro area and Switzerland with a Bayesian vector autoregressive structure (BVAR) and a factor model structure. He found evidence that the factor model structure performs satisfactorily.  

For neural network models, \cite{safi2016palestine} proposed a comparative study between artificial neural network models and time series forecasting models, finding results that indicate that the neural network model is superior to the ARIMA models and regression models for the quarterly Palestinian GDP in the period between 2014 to 2016. \cite{jahn2018artificial} applied an artificial neural network regression model to predict the annual GDP growth of fifteen industrialized countries, finding evidence that the model surpasses a similar linear model in the period from 1996 to 2016. \cite{tkacz2001canadian} investigates the predictive capacity of the artificial neural networks model for the forecasting of Canadian GDP growth, concluding that the model slightly outperforms the classic models for the short term and significantly exceeds for the long-term.  

Another accepted approach to GDP forecasting is macroeconomic projections based on leading indicators. \cite{garnitz2019gdp} applied this strategy to forecast GDP growth in forty-four countries, including Brazil. One of the results found indicates that the forecasts can be improved by adding World Economic Survey (WES) indicators of the three main trading partners by country.

The aim of this work is to investigate a suitable time series model to describe and forecast Brazilian GDP, also investigating the fit of these models to dynamics between periods of economic growth and recession. For this purpose, it is compared different classes of time series models. Thus, the chosen models were the Holt-Winters method, SARIMA, dynamic linear model, and the artificial neural network approach. In the literature, there are some applications regarding these models but no comparative studies were found using the models adopted in this work.

This work is organized as follows:  Section 2  describes the methodology. Section 3 presents the results and discussion, and, finally, the last section provides the main conclusions and some possibilities for future research.

\section{Methodology}
\label{sec:methodology}

To follow we outline the data and the empirical approach used to fitted and forecast the time series of Brazilian gross domestic product between the years 1996 and 2021, at 1995 prices. This section also defines the models that were investigated.

Care was also taken that the references used in the definition of models and metrics also correspond to studies and authors with wide use and quality proven by the academic community.

\subsection{Materials and Data}

The quality of the data used in empirical analysis is a fundamental element for the quality of the results. A factor that contributes to the empirical analysis of GDP is the vast documentation made available by government agencies. For that, we obtained the time series in the IBGE Automatic Recovery System \citep{ibge2020}. Statistical analyzes, as well as graphic representations, were built using open-source software \cite{r2020}.

The Table \ref{tab:description} shows the summary of statistics descriptions of data. The value of kurtosis coefficient indicates a platykurtic distribution for quarterly Brazilian GDP, i.e, the values of distribution is too flat. The skewness coefficient indicates that the distribution of data is highly skewed. We also accepted the hypothesis of stationary of data, according with Phillips-Perron unit root test for the first diff transformation.

\begin{table}[H]
\centering
\caption{Summary statistics for quarterly Brazilian GDP from 1Q1996 to 1Q2021, at 1995 prices. Note: *The p-value for Phillips-Perron unit root test for first diff transformation.}
\vspace{0.25cm}
\renewcommand{\arraystretch}{1.2}
\begin{tabular}{l|cccccccc}
\hline
\textbf{Description} & Sample size & Train size & Test Size & Kurtosis & Skewness & PP Test \\ \hline
\textbf{Value} & 101 & 88 & 13 & -1.570 & -0.212 & 0.01 \\\hline
\end{tabular}
\label{tab:description}
\end{table}

\subsection{Gross Domestic Product}

\begin{figure}[H]
	\centering
	\includegraphics[scale=0.47]{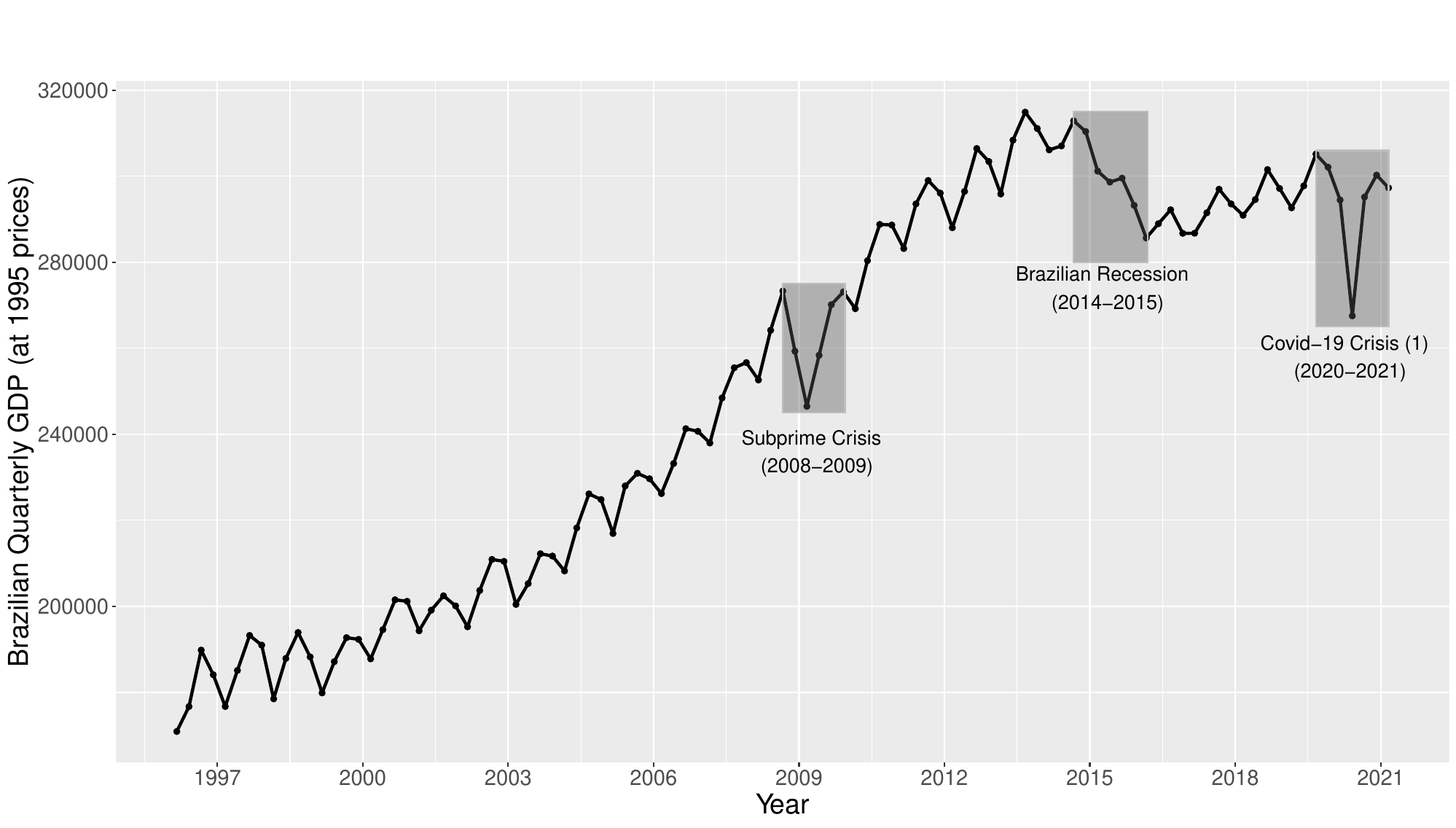}
	\caption{Time series of quarterly Brazilian GDP from 1Q1996 to 1Q2021, at 1995 prices. Authors elaboration.}
	\label{fig:gdp_plot}
\end{figure}

Several factors can affect the behavior of gross domestic product, and the economic/financial crisis are one of the most relevant. In the Figure \ref{fig:gdp_plot} we observe the time series of Brazilian quarterly GDP from 1Q1996 to 1Q2021. In this period, three events (or period of crisis) has highlight: the Subprime crisis (2008-2009), an external shock caused through a financial crisis in United States financial market that spread to the global market, including Brazil; the second event is the Brazilian recession period started in second quarter of 2014 with repercussions until fourth quarter of 2017, follow by a period of stagnation; and the third event is the Covid-19 crisis, a pandemic started in Wuhan, China and first reported in beginning of 2020 [\citep{wu2020},\citep{zhou2020}]. 

The \cite{SNA2008} says that GDP derives from the concept of value added. Therefore, GDP is the sum of gross value added of all resident producer units plus that part of taxes on products, fewer subsidies on products. GDP is also equal to the sum of finalizes of goods and services measured at purchasers’ prices, less the value of imports goods and services. And GDP is too equal the sum of primary incomes distributed by resident producer units.

According to \cite{feijo2013} GDP can be calculated in three different ways, but are part of the Accounting Identity (\textit{Production = Income = Expenditure}), guiding National Accounts. The perspective of production is calculated by sum the added values of economic activities plus taxes, net of subsidies, on products. That is,

\begin{equation}
Y_{production} = GVA-IC+(T-Sub),
\end{equation}

\noindent where \textit{GVA} it is  gross value added, \textit{IC} is the intermediate consumption, \textit{T} are taxes on products and \textit{Sub} are subsidies on products.

The income perspective is obtained by adding the remunerations of factors of production. Labor is remunerated by wages, loan capital is remunerated by interest, venture capital is remunerated by profit, and ownership of production goods ("land") is remunerated by rent. That is,

\begin{equation}
Y_{income} = W+GOS+(T-Sub),
\end{equation}

\noindent where \textit{W} are wages, \textit{GOS} are gross operating surplus (sum of interest, profit e rent), \textit{T} are taxes on products and \textit{Sub} are subsidies on products.

The time series constructed in this work was built from the perspective of expenditure. It is calculated by the sum of household consumption, investment, government spending and net exports. That is,

\begin{equation}
Y_{expenditure} = C+G+I+(NE),
\end{equation}

\noindent where \textit{C} it is household consumption, \textit{I} it is investment (gross fixed capital formation plus stock variation), \textit{G} it is government consumption, and \textit{NE} it is net exports (Exports less Imports).

\subsection{Holt-Winters Method}

As described in \cite{cowpertwait2009}, the Holt-Winters method was proposed by \cite{holt1957} and \cite{winters1960}, using exponentially weighted moving averages to update those needed for seasonal adjustment of the mean (trend) and seasonality.

The method has two variations with four equations: one forecast equation and three smoothing equations. \cite{hyndman2018forecasting} describes that in the additive method, the seasonal component is defined in absolute terms on the scale of the observed series. In the level equation, the series is seasonally adjusted by subtracting the seasonal component. Within each year, the seasonal component sum up to approximately zero. With the multiplicative method, the seasonal component is defined in percentage terms and the series is seasonally adjusted by dividing through by the seasonal component. Within each year the seasonal component will sum up to approximately $m$.

The additive method equations is describe as following,  

\begin{equation}
\begin{array}{l}
  \hat{y}_{t+h|t} = \ell_{t} + hb_{t} + s_{t+h-m(k+1)}, \\
  \ell_{t} = \alpha(y_{t} - s_{t-m}) + (1 - \alpha)(\ell_{t-1} + b_{t-1}), \\
  b_{t} = \beta(\ell_{t} - \ell_{t-1}) + (1 - \beta)b_{t-1}, \\
  s_{t} = \gamma (y_{t}-\ell_{t-1}-b_{t-1}) + (1-\gamma)s_{t-m},
\end{array}
\end{equation}

\noindent where $\hat{y}_{t+h|t}$ is the forecast equation. The $\ell_{t}$, $b_{t}$ and $s_{t}$ are respectively level, trend and seasonality equations, with corresponding smoothing parameters $\alpha$, $\beta$ and $\gamma$. The parameter $m$ denotes the frequency of seasonality, and for quarterly data $m=4$. Finally, $k$ is the integer part of $(\frac{h-1}{m})$ which ensures that the estimates of the seasonal indices used for forecasting come from the final year of the sample.

For the multiplicative method the same equations $\ell_{t}$, $b_{t}$ and $s_{t}$ are defined. But the change in structure occurs because instead of sum the equations in $\hat{y}_{t+h|t}$ an operation is performed to multiply the sum of the level and trend equations by the seasonality equation.

\subsection{SARIMA}

Box \& Jenkins models determine the proper stochastic process to represent a given time series by passing white noise through a linear filter \citep{morettin2018}. The model used was SARIMA, seeking to incorporate the seasonality component that is present in the data under analysis. 

The SARIMA of order $(p,q,d) \times (P,Q,D)_{s}$ is defined by,

\begin{equation}
\phi(B)\Phi(B^s)\nabla^d\nabla^D_{s}Y_{t} = \theta(B)\Theta(B^s)\alpha_{t},
\end{equation}

\noindent where $\theta(B)$ is the moving average operator of \textit{q} order, $\phi(B)$ is the autoregressive operator of \textit{p} order, $\Phi(B^s)$ is the seasonal autoregressive operator of \textit{P} order, $\Theta(B^s)$ is the seasonal moving average operator of \textit{Q} order, $\nabla^d$ is the simple difference operator, $\nabla^D_{s}$ is the seasonal difference operator and $\alpha_{t}$ is the noise.

\subsection{Artificial Neural Networks}

The artificial neural networks model seeks to model the relationship between a set of input signals and an output signal. We apply the multilayer perceptron (MLP) structure. By definition, the multilayer perceptron, or feedforward deep network, is a mathematical function mapping a sort of inputs values to output values. Following the contribution of \cite{goodfellow2016deep}, the main objective of a feedforward deep network is to approximate any function $f^{*}$, defining a mapping $y=f(x; \theta)$ and learns the value of the parameter $\theta$ that makes the better function approximation. 

We can describe a feedforward neural network through a hidden layer and a layer of lagged inputs, being a useful approach for forecasting univariate time series. When lagged values of the time series are uses as inputs to a feedforward neural network, this process is called neural network autoregression or NNAR model \citep{hyndman2018forecasting}. We can consider the relationship between the output and the inputs of neural network autoregression as 

\begin{equation}
    y_{t}=w_{0}+\sum_{j=1}^{h} w_{i} \cdot g\left(w_{0, j}+\sum_{i=1}^{n} w_{i, j} \cdot y_{t-j}\right)+\varepsilon
\end{equation}

\noindent where $y_{t}$ is the output, $\left({y}_{{t}-1},\ldots, y_{{tp}}\right)$ are the inputs, the model parameters (weights) are $w_{ij}(i=1,2,\dots,n; j=1,2,\dots,h)$, and $w_{j}(j=1,2,\dots,h)$. 

The usually activation function used is a sigmoid function, given by

\begin{equation}
    sig(x)=\frac{1}{1+e^{-x}}
\end{equation}

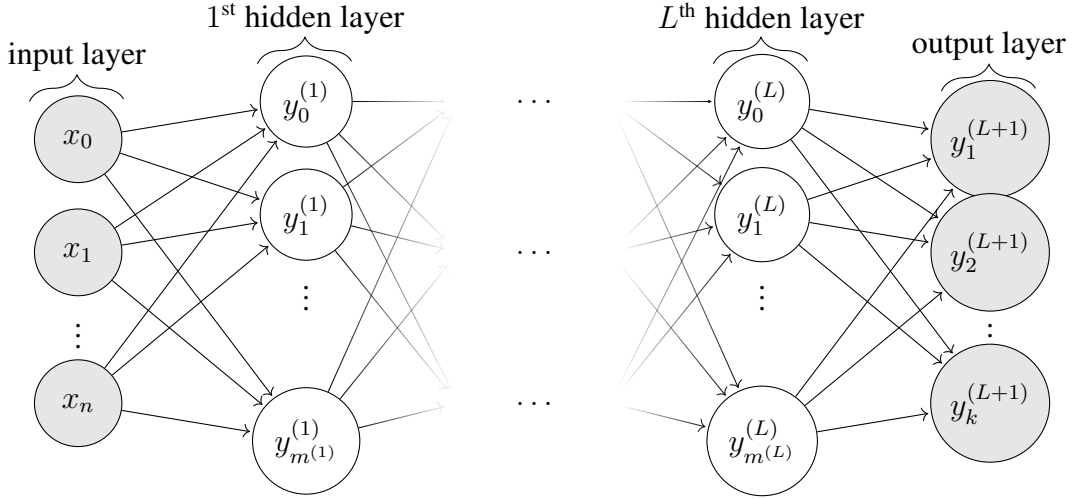
\begin{figure}[H]
	\centering
	\begin{tikzpicture}[shorten >=1pt]
		\tikzstyle{unit}=[draw,shape=circle,minimum size=1.15cm,fill=gray!20]
		\tikzstyle{hidden}=[draw,shape=circle,minimum size=1.15cm]

		\node[unit](x0) at (0,3.5){$x_0$};
		\node[unit](x1) at (0,2){$x_1$};
		\node at (0,1){\vdots};
		\node[unit](xd) at (0,0){$x_n$};

		\node[hidden](h10) at (3,4){$y_0^{(1)}$};
		\node[hidden](h11) at (3,2.5){$y_1^{(1)}$};
		\node at (3,1.5){\vdots};
		\node[hidden](h1m) at (3,-0.5){$y_{m^{(1)}}^{(1)}$};

		\node(h22) at (5,0){};
		\node(h21) at (5,2){};
		\node(h20) at (5,4){};
		
		\node(d3) at (6,0){$\ldots$};
		\node(d2) at (6,2){$\ldots$};
		\node(d1) at (6,4){$\ldots$};

		\node(hL12) at (7,0){};
		\node(hL11) at (7,2){};
		\node(hL10) at (7,4){};
		
		\node[hidden](hL0) at (9,4){$y_0^{(L)}$};
		\node[hidden](hL1) at (9,2.5){$y_1^{(L)}$};
		\node at (9,1.5){\vdots};
		\node[hidden](hLm) at (9,-0.5){$y_{m^{(L)}}^{(L)}$};

		\node[unit](y1) at (12,3.5){$y_1^{(L+1)}$};
		\node[unit](y2) at (12,2){$y_2^{(L+1)}$};
		\node at (12,1){\vdots};	
		\node[unit](yc) at (12,0){$y_k^{(L+1)}$};
        
        \draw[->] (x0) -- (h10);
		\draw[->] (x0) -- (h11);
		\draw[->] (x0) -- (h1m);

		\draw[->] (x1) -- (h10);
		\draw[->] (x1) -- (h11);
		\draw[->] (x1) -- (h1m);

		\draw[->] (xd) -- (h10);
		\draw[->] (xd) -- (h11);
		\draw[->] (xd) -- (h1m);

		\draw[->] (hL0) -- (y1);
		\draw[->] (hL0) -- (yc);
		\draw[->] (hL0) -- (y2);

		\draw[->] (hL1) -- (y1);
		\draw[->] (hL1) -- (yc);
		\draw[->] (hL1) -- (y2);

		\draw[->] (hLm) -- (y1);
		\draw[->] (hLm) -- (y2);
		\draw[->] (hLm) -- (yc);
        
        \draw[->,path fading=east] (h10) -- (h20);
		\draw[->,path fading=east] (h10) -- (h21);
		\draw[->,path fading=east] (h10) -- (h22);
		
		\draw[->,path fading=east] (h11) -- (h20);
		\draw[->,path fading=east] (h11) -- (h21);
		\draw[->,path fading=east] (h11) -- (h22);
		
		\draw[->,path fading=east] (h1m) -- (h20);
		\draw[->,path fading=east] (h1m) -- (h21);
		\draw[->,path fading=east] (h1m) -- (h22);
		
		\draw[->,path fading=west] (hL10) -- (hL0);
		\draw[->,path fading=west] (hL11) -- (hL0);
		\draw[->,path fading=west] (hL12) -- (hL0);
		
		\draw[->,path fading=west] (hL10) -- (hL1);
		\draw[->,path fading=west] (hL11) -- (hL1);
		\draw[->,path fading=west] (hL12) -- (hL1);
		
		\draw[->,path fading=west] (hL10) -- (hLm);
		\draw[->,path fading=west] (hL11) -- (hLm);
		\draw[->,path fading=west] (hL12) -- (hLm);
		
		\draw [decorate,decoration={brace,amplitude=10pt},xshift=-4pt,yshift=0pt] (-0.5,4) -- (0.75,4) node [black,midway,yshift=+0.6cm]{input layer};
		\draw [decorate,decoration={brace,amplitude=10pt},xshift=-4pt,yshift=0pt] (2.5,4.5) -- (3.75,4.5) node [black,midway,yshift=+0.6cm]{$1^{\text{st}}$ hidden layer};
		\draw [decorate,decoration={brace,amplitude=10pt},xshift=-4pt,yshift=0pt] (8.5,4.5) -- (9.75,4.5) node [black,midway,yshift=+0.6cm]{$L^{\text{th}}$ hidden layer};
		\draw [decorate,decoration={brace,amplitude=10pt},xshift=-4pt,yshift=4pt] (11.5,4) -- (12.75,4) node [black,midway,yshift=+0.6cm]{output layer};
	\end{tikzpicture}
	\caption[Network graph for a $(L+1)$-layer perceptron.]{Network graph of a $(L+1)$-layer perceptron with $n$ input units and $k$ output units. The $l^{\text{th}}$ hidden layer contains $m^{(l)}$ hidden units.}
	\label{fig:multilayer-perceptron}
\end{figure}

In Figure \ref{fig:multilayer-perceptron} is possible to observe the graph representation of an MLP as describes above, with $n$ inputs in input layer, $L$ hidden layers with $m^(L)$ hidden units, and $k$ outputs in output layer. 

\subsection{Dynamic Linear Model}

Dynamic linear models are an important class of state-space models. Broadly used in the last decades, they have a high degree of efficiency for the analysis and forecast of time series, providing flexibility and applicability through an elegant and robust probabilistic apparatus. 

The estimation and inference challenges are solved by recursive algorithms, which follow the Bayesian approach, calculating conditional distributions of quantities of interest given the observed information. Considering a series affected by time, through dynamic and random deformations, they associate seasonal or regressive components.

In this work were used contributions from \cite{west1997}, \cite{laine2019}, \cite{petris2009} and \cite{petris2010}. For each time $t$, the general univariate DLM is defined by a observational equation,

\begin{equation}
Y_{t} = F_{t}\theta_t + v_{t}, \qquad v_{t} \sim N_{m}(0,V_{t}),
\end{equation}
\noindent a system equation

\begin{equation}
\theta_{t} = G_{t}\theta_{t-1}+w_{t}, \qquad w_{t} \sim N_{p}(0,W_{t})
\end{equation}
\noindent and initial information given by

\begin{equation}
(\theta_{0}|D_{0})\sim N(m_{0}, C_{0}),
\end{equation}
\noindent where $F_{t}$ e $G_{t}$ are known matrices; $v_{t}$ and $w_{t}$ are two sequences of independent noises, with average zero and known covariance matrices $V_{t}$ and $W_{t}$ respectively. $D_{t}$ is the current information set; $m_{0}$ and $C_{0}$ contains relevant information about the future, according usual statistical sense, given $D_{t}$, ($m_{t}$, $C_{t}$) is sufficient for $(Y_{t+1}, \theta_{t+1}, \dots, Y_{t+k}, \theta_{t+k})$.

To take into account growth and seasonality, it is defined $\theta_t=(\mu_t,\beta_t,\gamma_t, \gamma_{t-1},\gamma_{t-2})$, where $\mu_{t}$ is the current level, $\beta_{t}$ is the slope of the trend, $\gamma_{t},~\gamma_{t-1}~\text{and}~\gamma_{t-2}$ are the seasonal components.

\subsection{Metrics}
\label{subsec:metrics}

The selection of most suitable forecasting model was made through the contributions of \cite{hyndman2006}, \cite{armstrong2001}, \cite{morettin2018} and \cite{ahlburg1984theil} using the mean absolute percentage error (MAPE).

The MAPE precision metric has the advantage of being scale-independent, and so are frequently used to compare forecast performance across different data sets. The metric is defined as 

\begin{equation}
    MAPE(y_{t},\hat{y}_{t+h}) = 100\times|\frac{y_{t} - \hat{y}_{t+h|t}}{y_{t}}|,
\end{equation}

\noindent where $y_{t}$ is the observed value $y$ in time $t$ and $\hat{y}_{t+h}$ is the predicted value $\hat{y}_{t}$ with $h$ steps ahead.

\subsection{Empirical Strategy}

\begin{figure}[H]
    \centering
    \includegraphics[scale=0.7]{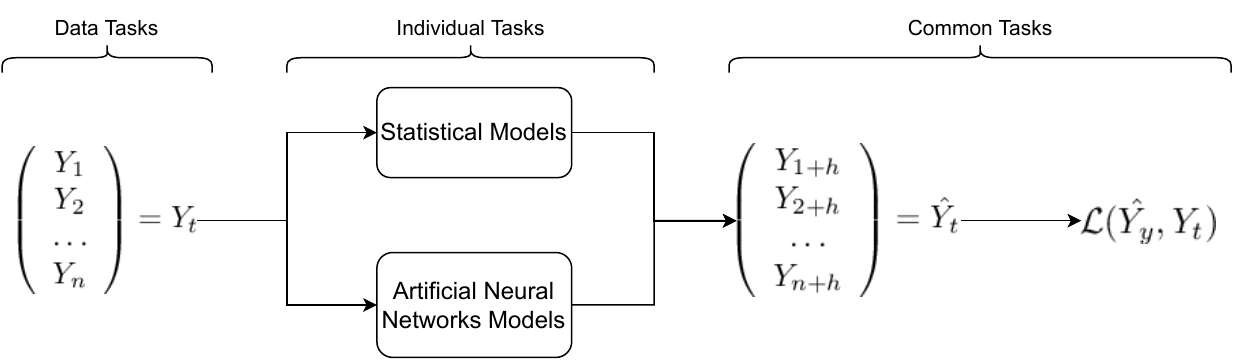}
    \caption{Flowchart of empirical strategy adopted in study. Authors elaboration.}
    \label{fig:workflow}
\end{figure}

The empirical strategy and forecast workflow (Figure \ref{fig:workflow}) of this study taking into account three basic steps: the data tasks, the individual tasks, and the common tasks. In data task step we first extract the variables of quarterly Brazilian gross domestic product, by expenditure side, from IBGE Automatic Recovery System (SIDRA). Next, we initialize the organization of data: (i) compute GDP as $Y = C+G+I+(NE)$; (ii) convert $Y$ to time series format ($Y_{t}$); (iii) split $Y_{t}$ in train and test sets.
    
In individual tasks step we define the structure uses in the three classes of models that we consider (classical, state-space, and artificial neural networks). For classic models first we compute the sum of squared errors (SSE) for additive and multiplicative methods. The method with lower SSE was considered for analysis; next, we apply the selection algorithm of SARIMA structure (described in Algorithm \ref{alg:sarima_select}). In state-space approach we use the dynamic linear model with variance parameters estimated on Monte Carlo Markov Chain method, through the Gibbs Sampler. Thus, the structure of matrices of dynamic linear model was constructed based on Gibbs Sampler results. And in the artificial neural networks we consider the multilayer perceptron (MLP) with different number of layers and the neural network autoregression (NNAR).

For common tasks step we compute the predictions ($\hat{Y_{t}}$) of each model applied to the dataset under analysis and then calculate the prediction error for the training and testing sets. Then the graphical representations and comparison tables of the metrics are created.

\section{Results and Analysis}

This section presents the results obtained using the Holt-Winters additive method, SARIMA, dynamic linear models, neural network autoregression, and multilayer perceptron to fit the data of interest. For each model, it was plotted the mean absolute error for a forecast horizon ($h$) varying $h=1,\dots,13$. 

\subsection{Holt-Winters Method}

\begin{figure}[H]
	\centering
	\includegraphics[scale=0.35]{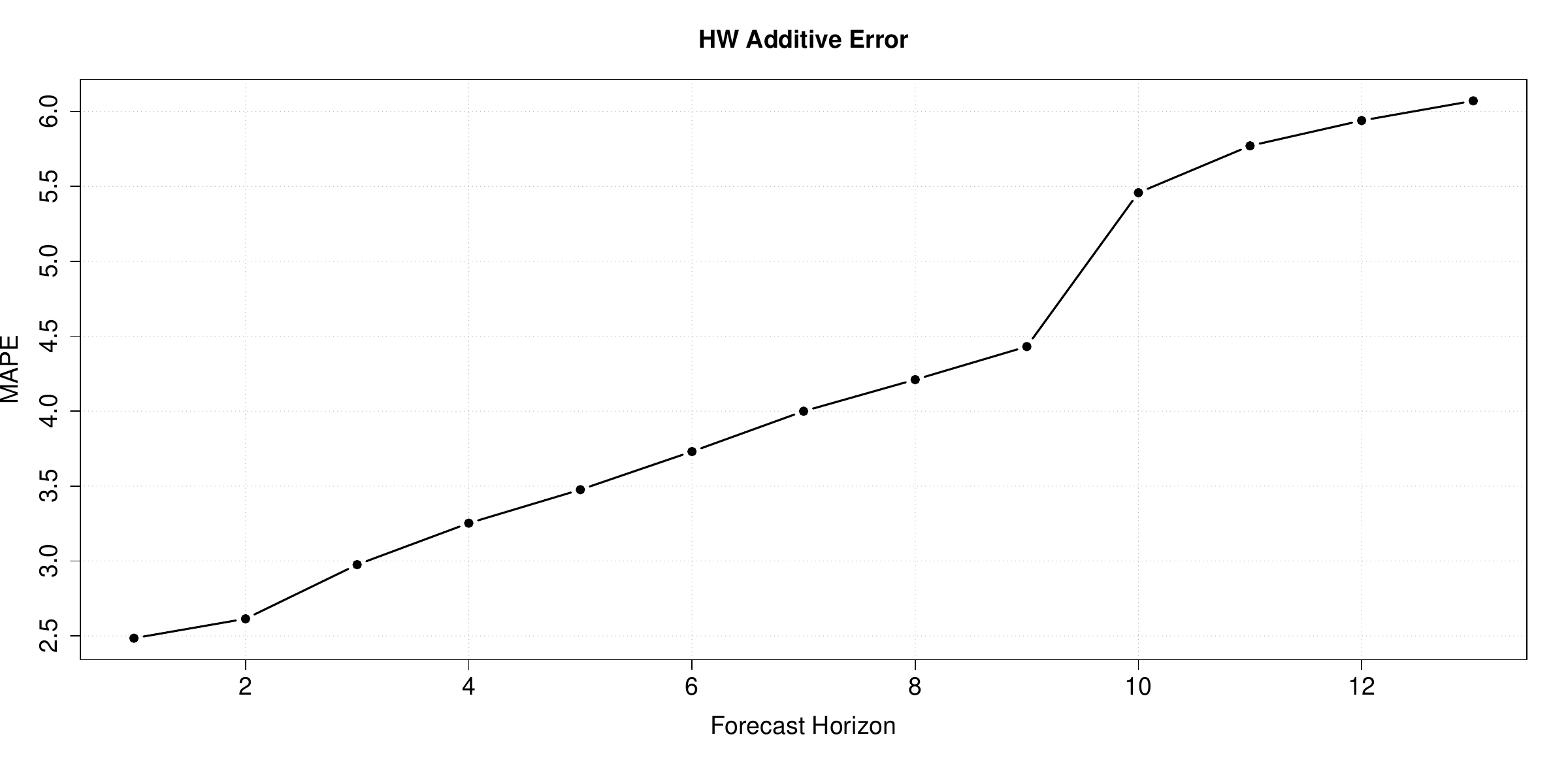}
	\caption{MAPE results for Holt-Winters Additive applied to Brazilian GDP, at 1995 prices.}
	\label{fig:hw_plot1}
\end{figure}

Compared to the multiplicative Holt-Winters method, the additive formulation  was considered the most appropriate, taking into account the sum of squared errors. Figure \ref{fig:hw_plot1} shows the evolution trajectory of MAPE through the Holt-Winters Method. It can be seen that MAPE growth increases linearly with the steps of the forecast horizon.

\subsection{SARIMA}

\begin{figure}[H]
	\centering
	\includegraphics[scale=0.4]{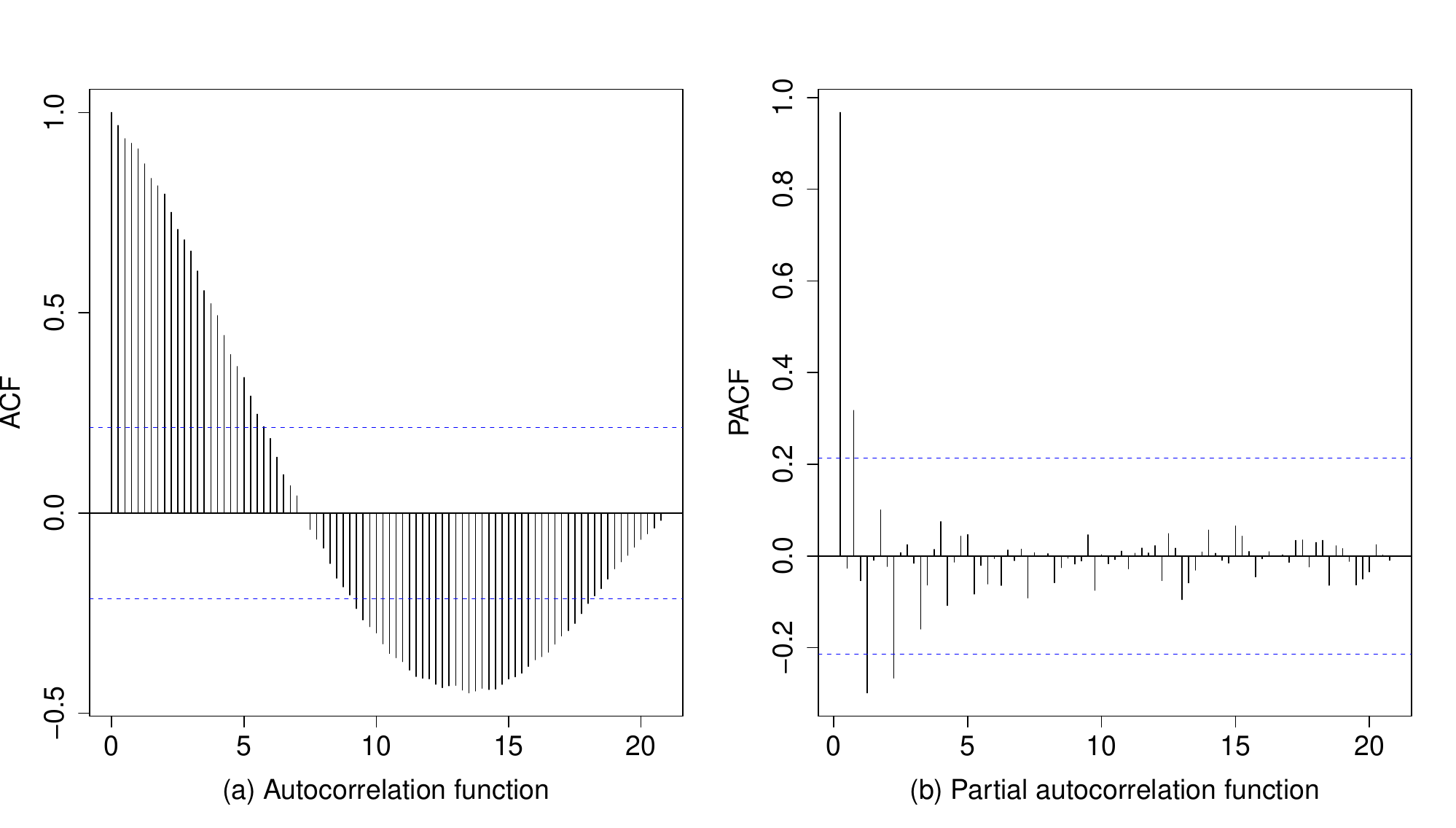}
	\caption{Autocorrelation function (a) and partial autocorrelation function (b) to the observed Brazilian quarterly GDP in the period from 1996 to 2019, at 1995 prices.}
	\label{fig:acf}
\end{figure}

To apply SARIMA model, the behavior of autocorrelation (ACF) and partial autocorrelation functions (PACF) were verified. In Figure \ref{fig:acf} (a), it is possible to see a slow decay rate of the autocorrelation function to zero. This behavior indicates the non-stationarity of the series, which needs to be differentiated in order to make it stationary.

\begin{figure}[H]
	\centering
	\includegraphics[scale=0.4]{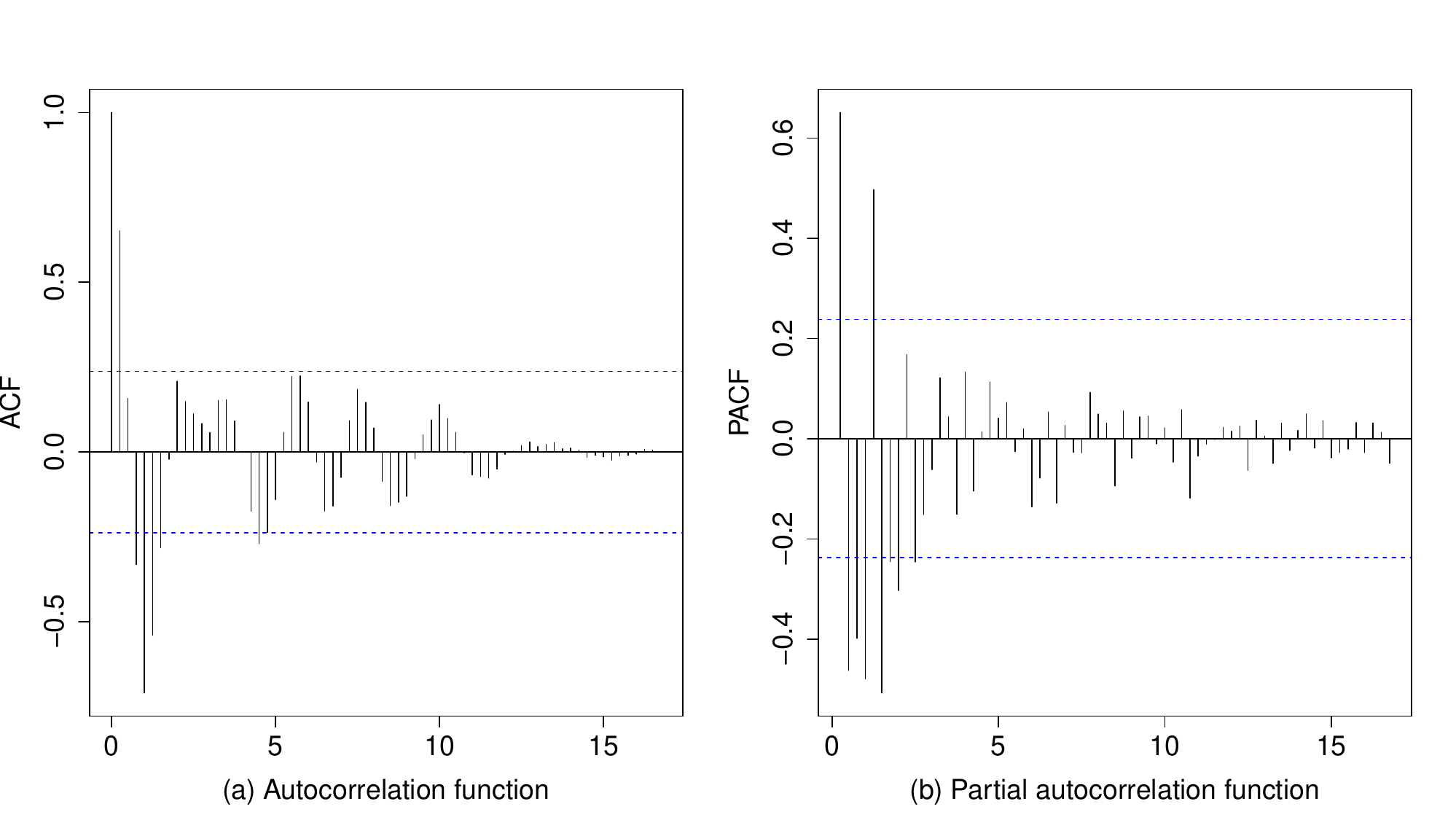}
	\caption{Autocorrelation function (a) and partial autocorrelation function (b) to the differentiated series of Brazilian quarterly GDP in the period from 1996 to 2019, at 1995 prices.}
	\label{fig:acf_diff}
\end{figure}

Figure \ref{fig:acf_diff} (a) shows the autocorrelation function of the differentiated series with an exponential decay in the lags multiples of 4, indicating a possible series stationarity. Through the Phillips-Perron test (Dickey-Fuller $ Z_{\alpha}$ = -62.816; p-value = 0.01), the alternative hypothesis of stationarity of the differentiated series was accepted at a significance level of 1\%. 

We used an algorithm (Algorithm \ref{alg:sarima_select}) to generate sixteen SARIMA models following the principle of parsimony. From the generated models, the structure with the best results was the SARIMA $(1,1,0)\times(1,1,1)_{4}$, with metrics: MAPE (0.9823); RMSE (3216.698); and the Ljung-Box test (p-value = 0.997), showing that residuals are independently distributed. However, other structures presented good results considering the selection algorithm results.

\vspace{0.5cm}
\begin{algorithm}[H]
\SetAlgoLined
Initial Parameters: $DD = 1, d=1, per=4$\;
For each instance: $p,q,i,j=1,2$, if the principle of parsimony is valid: p+d+q+i+DD+j $\leq$10\;
    \hspace{0.3cm} 0. run the SARIMA model with training set\;
    \hspace{0.3cm} 1. generate the p-value of Ljung-Box test of residuals\;
    \hspace{2cm} $H_{0}:$ the data are independently distributed\\
    \hspace{2cm} $H_{a}:$ the data are not independently distributed\\
    \hspace{2cm} $Q=n(n+2)\sum_{k=1}^{h}\frac{\hat{\rho}^{2}_{k}}{n-k}$\;
    \hspace{0.3cm} where n denotes the sample size, $\hat{\rho}^{2}_{k}$ is the sample autocorrelation at lag $k$, $h$ is the number of tested lags\;
    \hspace{0.3cm} 2. compute the mean absolute percentage error \\
    \hspace{2cm} $\frac{1}{n}\sum\frac{|y_{t} - \hat{y}_{t}|}{|y_{t}|}\times 100$\;
    \hspace{0.3cm} 3. compute the root of mean squared error\;
    \hspace{2cm} $\sqrt{\frac{(y_{t}-\hat{y_{r})^{2}}}{n}}$\;
    \hspace{0.3cm} 4. aggregate the results\;
    END
    
\caption{The selection algorithm for SARIMA structure by principle of parsimony.}
\label{alg:sarima_select}
\end{algorithm}
\vspace{0.5cm}

\begin{figure}[H]
	\centering
	\includegraphics[scale=0.35]{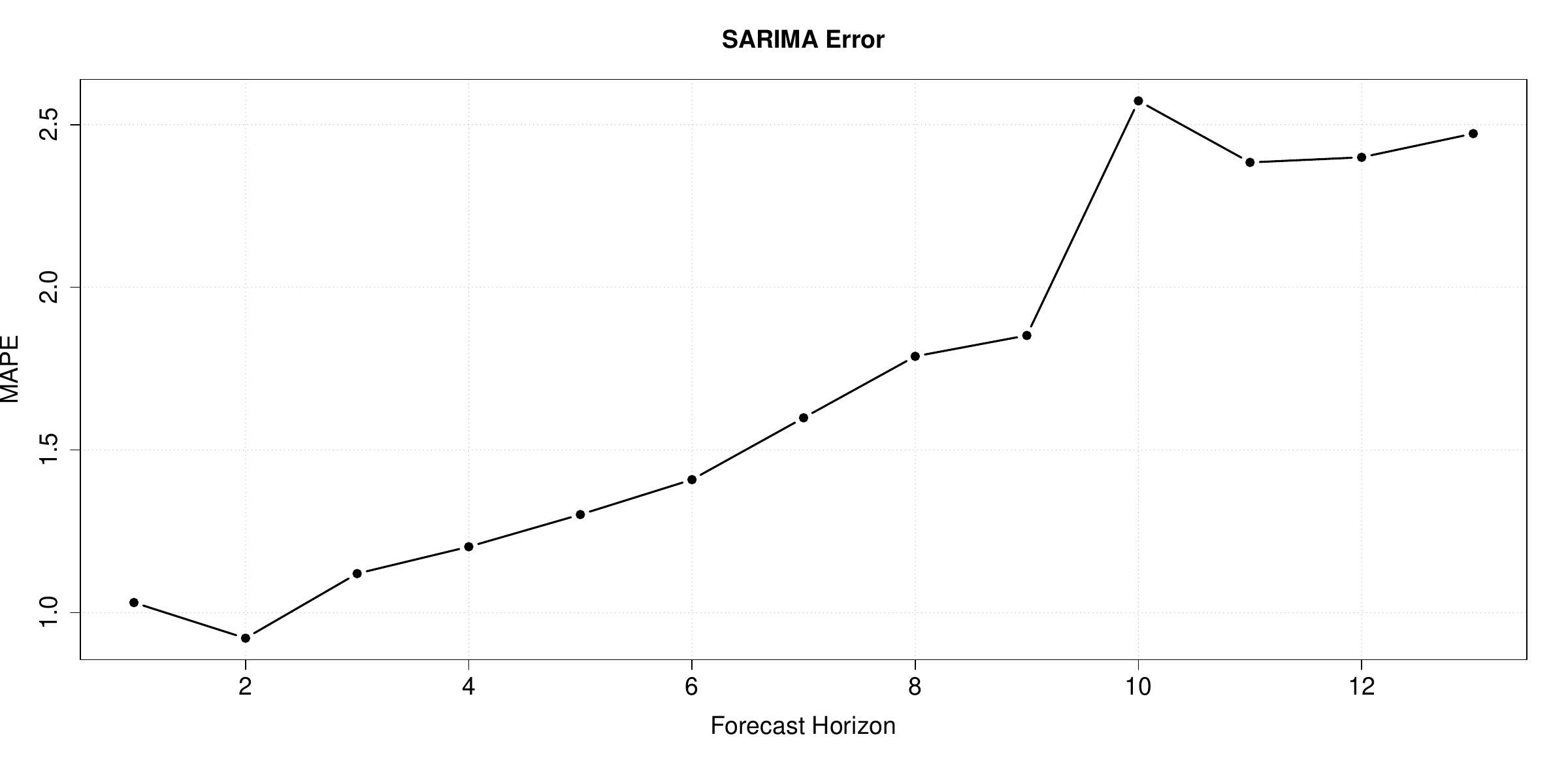}
	\caption{MAPE results for SARIMA applied to Brazilian GDP, at 1995 prices.}
	\label{fig:sarima_plot}
\end{figure}

Figure \ref{fig:sarima_plot} shows the evolution trajectory of MAPE through the SARIMA model. The MAPE is below 2\% up to the forecast horizon of 9 steps ahead.

\subsection{Artificial Neural Networks}

In this study, we consider the neural network autoregression model, NNAR$(p,P,k)_{m}$. The model takes into account seasonality with input layer as lags $(y_{t-1}, y_{t-2},\dots, y_{tp}, y_{tm}$, $y_{t-2m}, y_{t- Pm})$ with a hidden layer with $k$ nodes.

\begin{figure}[H]
	\centering
	\includegraphics[scale=0.35]{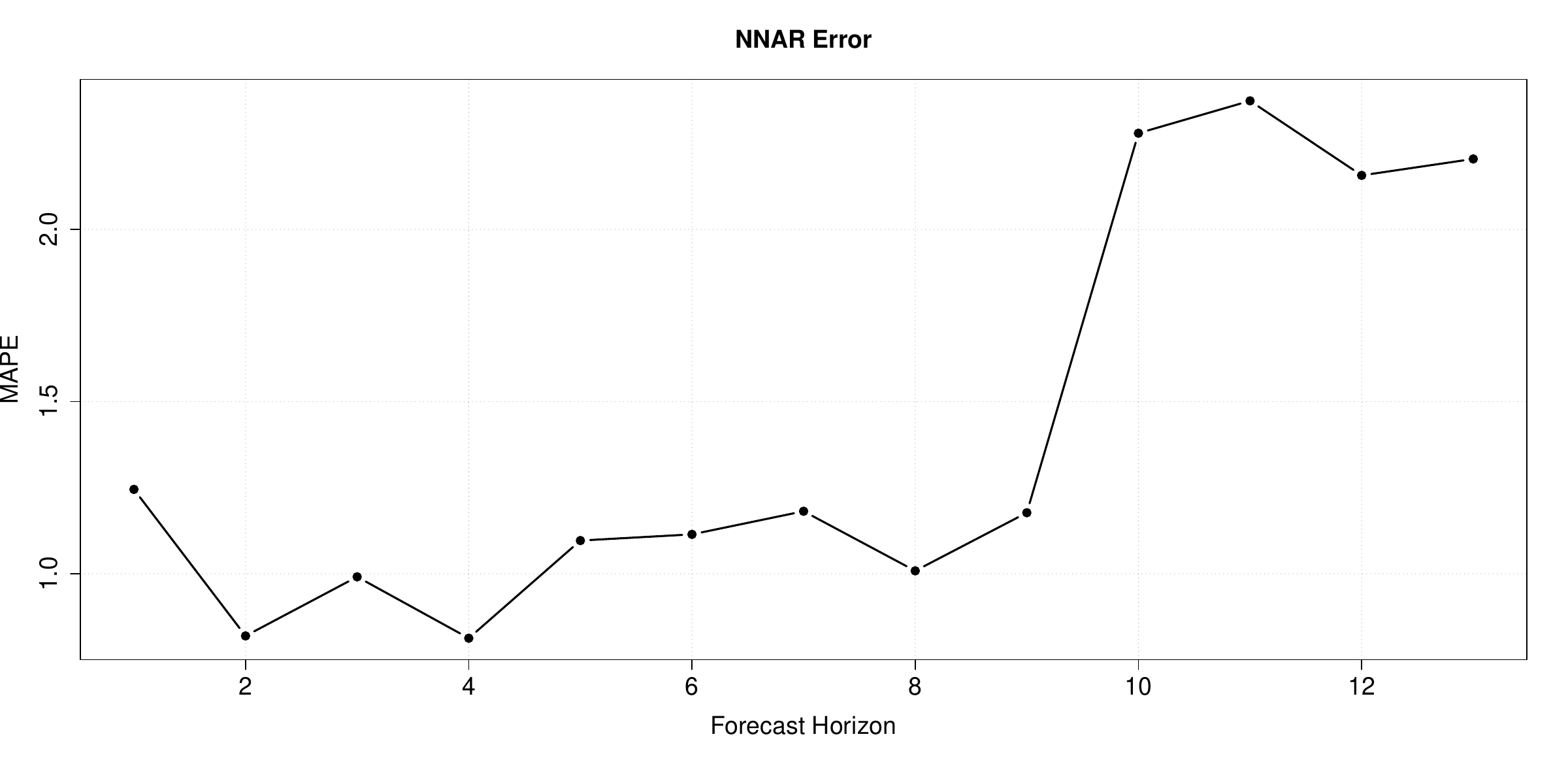}
	\caption{MAPE results for Neural Networks Autoregression applied to Brazilian GDP, at 1995 prices.}
	\label{fig:nnar_plot}
\end{figure}

\begin{figure}[H]
	\centering
	\includegraphics[scale=0.35]{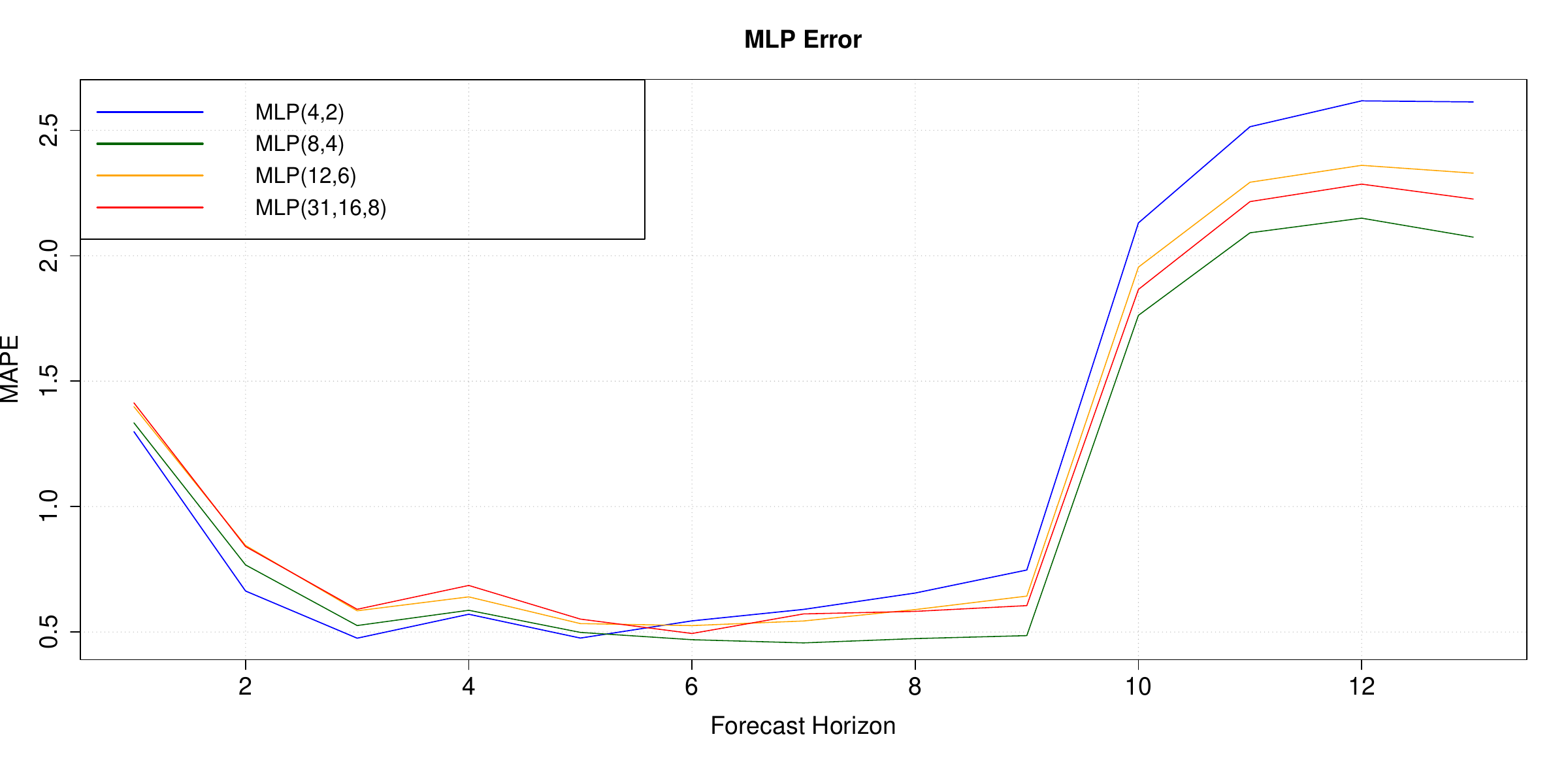}
	\caption{MAPE results for Multilayer Perceptron applied to Brazilian GDP, at 1995 prices.}
	\label{fig:mlp_plot}
\end{figure}

Figure \ref{fig:nnar_plot} shows the trajectory of MAPE through NNAR model. It can be seen in this figure that the MAPE is less than 1\% up to the forecast horizon of 9 steps ahead. This is the same behavior (Figure \ref{fig:mlp_plot}) observed for MLP models\footnote{MLP1: 2 hidden layers with 4 and 2 nodes; MLP2: 2 hidden layers with 8 and 4 nodes; MLP3: 2 hidden layers with 12 and 6 nodes; and MLP4: 3 hidden layers with 31, 16, and 2 nodes;}.

\subsection{Dynamic Linear Model}

In this work, the dynamic regression matrix $F_{t}$ and the evolution matrix $G_{t}$ of the model are

\begin{equation}
\begin{array}{ll}
{F_t} =    
\begin{bmatrix}
1 & 0 & 1 & 0 & 0
\end{bmatrix}
\qquad\mbox{and}\qquad
{G_t} = 
\begin{bmatrix}
1 & 1 & 0 & 0 & 0 \\
0 & 1 & 0 & 0 & 0 \\
0 & 0 & -1 & -1 & -1 \\
0 & 0 & 1 & 0 & 0 \\
0 & 0 & 0 & 1 & 0
\end{bmatrix}
\end{array}
.
\end{equation}

For the study, it was assumed the observational variance $V_t=\sigma^2$, and the covariance matrix of the system $W_t$ is a diagonal matrix introduced by  $W_t = diag(\sigma_{\mu}^2, \sigma_{\beta}^2, \sigma_{\gamma}^2,0,0)$. These unknown variances were also estimated using Bayesian inference. Thus, to complete the specification of the model, it was assumed independent inverse gamma priors distributions with means $a, a_{\theta_1}, a_{\theta_2}, a_{\theta_3} ~\text{and  variances}~ b, b_{\theta_1}, b_{\theta_2}, b_{\theta_3},$ respectively, fixed in known values.

Therefore, by using the unobservable states as latent variables, a Gibbs sampler can be run on the basis of the following full conditional densities:

\begin{equation}\displaystyle
\begin{array}{l}
  \sigma^2 \sim IG\left(\frac{a^2}{b}+\frac{n}{2}, \frac{a}{b}+\frac{1}{2}SS_y\right),\\ \vspace{0.2cm}
  \sigma_{\mu}^2 \sim IG\left(\frac{a_{\theta_1}^2}{b_{\theta_1}}+\frac{n}{2}, \frac{a_{\theta_1}}{b_{\theta_1}}+\frac{1}{2}SS_{\theta_1}\right),\\ \vspace{0.2cm}
   \sigma_{\beta}^2 \sim IG\left(\frac{a_{\theta_2}^2}{b_{\theta_2}}+\frac{n}{2}, \frac{a_{\theta_2}}{b_{\theta_2}}+\frac{1}{2}SS_{\theta_2}\right),\\ \vspace{0.2cm}
    \sigma_{\gamma}^2 \sim IG\left(\frac{a_{\theta,3}^2}{b_{\theta_3}}+\frac{n}{2}, \frac{a_{\theta_3}}{b_{\theta_3}}+\frac{1}{2}SS_{\theta_3}\right),\\
\end{array}
\end{equation}
with $SS_y = \sum_{t=1}^n\left(y_t - F_t \theta_t\right)^2$ and $    SS_{\theta_i} = \sum_{t=1}^T \left(\theta_{t,i} - (G_t\theta_{t-1})_i\right)^2$, for $i=1,~2,~3$. The full conditional density of the states is a normal distribution and it is covered in the used dlm package \citep{petris2010}.\\

\begin{figure}[H]
	\centering
	\includegraphics[scale=0.35]{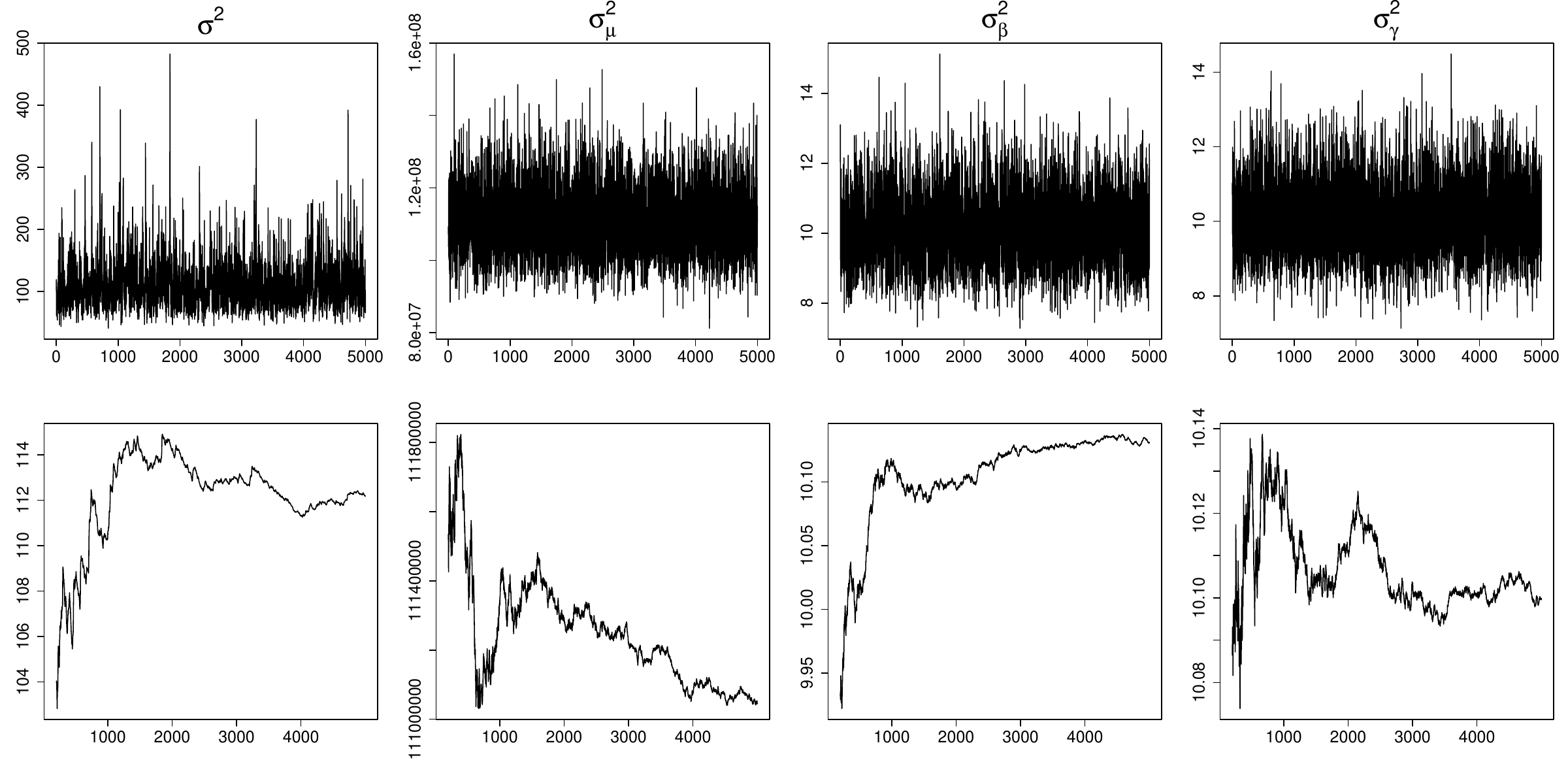}
	\caption{Trajectory  of simulated variances (top) and the ergodic means (bottom).}
	\label{fig:dlm_variance}
\end{figure}

From the Gibbs sampler, 5000 iterations were generated for each parameter, model variances, out of which the 1000 initial iterations were considered as burn-in period and discarded. Hence, the remaining iterations were used to compose the posterior samples of the estimated variances. Posterior estimates of the four unknown variances, from the Gibbs sampler output, can be seen in Figure \ref{fig:dlm_variance}. 

Figure \ref{fig:dlm_variance} shows the trajectory of MAPE. It can be seen in this figure that the MAPE is less than 2\% for up to 9 steps ahead forecasts horizons.

\begin{figure}[H]
	\centering
	\includegraphics[scale=0.35]{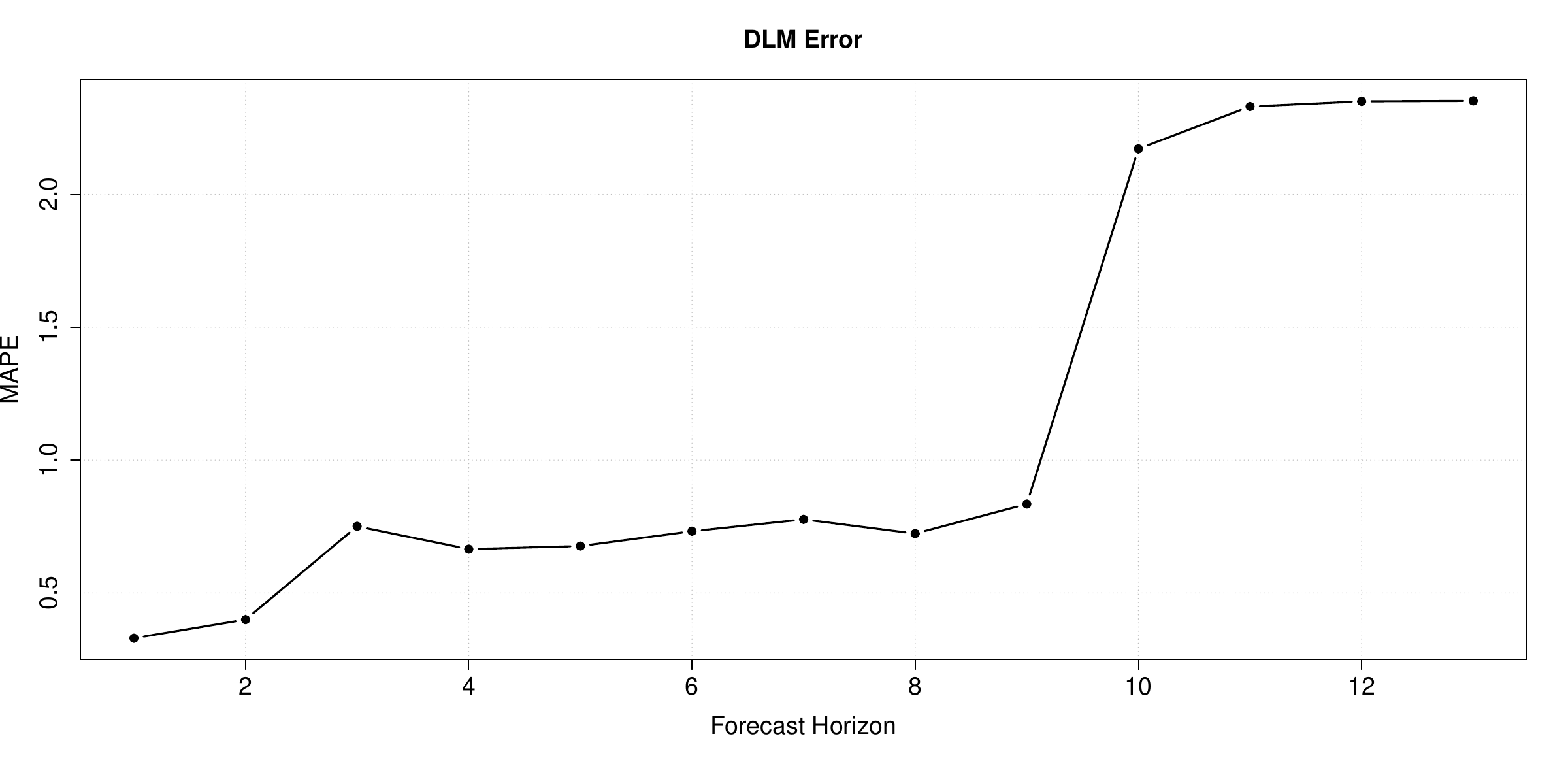}
	\caption{MAPE results for Dynamic Linear Model applied to Brazilian GDP, at 1995 prices.}
	\label{fig:hw_plot}
\end{figure}

\subsection{Model Comparison}

The metrics were used to assess the goodness of fit of models to the Brazilian quarterly GDP data between the years 1996 and 2016, at 1995 prices, and their results are shown in Table 1. It is observed that the better results were given through the multilayer perceptron models, this model being one that best fits the series of Brazilian GDP, at 1995 prices, for having achieved the lowest values in all metrics for fitted and forecast values.

\begin{table}[H]
	\centering
	\def\arraystretch{1}
	\caption{Fitted MAPE result for models applied to Brazilian quarterly GDP, at 1995 prices.}
	\vspace{0.5cm}
	\bgroup
	\begin{tabular}{l l c }
		\toprule
		& Model &  MAPE \\ 
        \hline
        1 & Holt-Winters & 1.48 \\ 
        2 & SARIMA & 0.97 \\ 
        3 & NNAR & 1.48 \\ 
        4 & DLM & 0.77 \\ 
        5 & MLP1 & 0.90 \\ 
        6 & MLP2 & 0.82 \\ 
        7 & MLP3 & 0.79 \\ 
        8 & MLP4 & \textbf{0.61} \\
		\bottomrule
	\end{tabular}	
	\egroup
	\label{tab:fitted}
\end{table}

Table \ref{tab_comp} presents the MAPE for each forecast horizon yielded by proposed models.The dynamic linear model presented the best MAPE result for forecasts up to two steps ahead. But, the multilayer perceptons models were better for the other forecasting periods.

\begin{table}[H]
\centering 
\caption{Forecast MAPE result for models applied to Brazilian quarterly GDP, at 1995 prices.}
\vspace{0.25cm}
\renewcommand{\arraystretch}{1.2}
\resizebox{\textwidth}{!}{%
\begin{tabular}{llrrrrrrrrrrrrr}
  \hline
& Forecast Horizon & 1 & 2 & 3 & 4 & 5 & 6 & 7 & 8 & 9 & 10 & 11 & 12 & 13 \\ 
  \hline
  1 & Holt-Winters & 2.48 & 2.61 & 2.98 & 3.25 & 3.48 & 3.73 & 4.00 & 4.21 & 4.43 & 5.46 & 5.77 & 5.94 & 6.07 \\ 
  2 & SARIMA & 1.03 & 0.92 & 1.12 & 1.20 & 1.30 & 1.41 & 1.60 & 1.79 & 1.85 & 2.57 & 2.38 & 2.40 & 2.47 \\ 
  3 & NNAR & 1.23 & 0.87 & 0.95 & 0.85 & 1.12 & 1.10 & 1.06 & 1.01 & 1.16 & 2.38 & 2.40 & 2.15 & 2.25 \\ 
  4 & DLM & \textbf{0.33} & \textbf{0.40} & 0.75 & 0.66 & 0.68 & 0.73 & 0.78 & 0.72 & 0.83 & 2.17 & 2.33 & 2.35 & 2.35 \\ 
  5 & MLP1 & 1.33 & 0.69 & \textbf{0.52} & 0.62 & 0.54 & 0.60 & 0.67 & 0.76 & 0.87 & 2.25 & 2.65 & 2.78 & 2.79 \\ 
  6 & MLP2 & 1.39 & 0.81 & 0.56 & 0.63 & 0.54 & 0.53 & 0.54 & 0.57 & 0.61 & 1.92 & 2.28 & 2.36 & 2.31 \\ 
  7 & MLP3 & 1.37 & 0.84 & 0.58 & \textbf{0.61} & \textbf{0.50} & \textbf{0.50} & 0.52 & 0.57 & 0.62 & 1.95 & 2.30 & 2.38 & 2.34 \\ 
  8 & MLP4 & 1.39 & 0.87 & 0.58 & \textbf{0.61} & 0.52 & \textbf{0.50} & \textbf{0.49} & \textbf{0.52} & \textbf{0.56} &\textbf{ 1.82} & \textbf{2.12} & \textbf{2.14} & \textbf{2.04} \\ 
   \hline
\end{tabular}
} \label{tab_comp}
\end{table}

\section{Conclusion}

Understanding GDP behavior is a topic of study and discussion by society and the academic community. In the present work, we proposed the application of the Holt-Winters additive method, SARIMA, dynamic linear model, and neural network models with interest in the forecast of Brazilian quarterly GDP, at 1995 prices. The data comprise the period between the first quarter of 1996 and the fourth quarter of 2019. 

The results of the study indicate that the proposed models have a satisfactory ability to adjust to the Brazilian quarterly GDP data.  The MAPE of fit of the models to the training set was less than 2\%. In addition, the models managed to capture the complex structure of the data involving the crises (peaks in the series) in the years 2009, 2015 and 2020. By the metric MAPE, it was found that the multilayer perceptron presented the best fit to data and efficient forecast performance. On the other hand, the dynamic linear model presented the second best fit result and the best predictive ability for the forecast horizon of two steps ahead.

We find evidence in this study that corroborates with the observed results of stagnation in the Brazilian economy after a crisis period started in the second quarter of 2014. Therefore, the multilayer perceptron model proved to be efficient for forecasting and fit GDP data even with economic shocks.

For future work, it would be interesting to compare the results obtained with forecasting models of the machine learning approach. For example, with Long Short-Term Memory (LSTM) networks. And detect the concept drifts present in the time series, looking for detect crisis periods.

\pagebreak

\bibliography{references}

\end{document}